\documentclass[11pt, a4paper]{article}
\usepackage[a4paper, margin=2.5cm]{geometry}
\usepackage[utf8]{inputenc}
\usepackage[T1]{fontenc}

\usepackage{amsmath, amssymb}
\usepackage{booktabs}
\usepackage[authoryear, round]{natbib}
\usepackage[colorlinks=true, linkcolor=blue, citecolor=blue, urlcolor=blue]{hyperref}
\usepackage{graphicx}

\usepackage{orcidlink}
\renewcommand{\orcidlink}[1]{\href{https://orcid.org/#1}{\includegraphics[width=10pt]{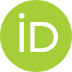}}} 

\title{\textbf{SIMULATION OF A JOVIAN PLANET UNDERGOING GAS ACCRETION}}
\author{\textbf{Fabian Menezes}$^1$ \orcidlink{0000-0002-4703-4027}}

\date{\small $^1$ Centro de R\'adio Astronomia e Astrof\'isica Mackenzie (CRAAM), Universidade Presbiteriana Mackenzie, Rua da Consola\c{c}\~ao, 930, S\~ao Paulo, SP, 01302-907, Brazil}

\begin{document}

\maketitle

{\noindent Contact: \url{menezes.astroph@gmail.com}}\\
{\noindent Published in \textit{Revista Mackenzie de Engenharia e Computação}, São Paulo, v. 18, n. 1, p. 67-82, 2018. Available at \url{https://editorarevistas.mackenzie.br/rmec/article/view/11018/7357}}

\begin{abstract}
To understand the process of planet formation in the Solar System or any other stellar system, we need to comprehend the interaction between forming planets and the protoplanetary disk. This work presents a study of the interaction of the protoplanetary disk with the protoplanet and the following types of orbital migration that can occur, and to verify this interaction, numerical simulations were performed using the magneto-hydrodynamics code FARGO3D. The researchers made five simulations with a 2-Jupiter-mass planet in the protoplanetary disk. The only parameter varied was the initial radial position with the aim of verifying the influence of this parameter on the behavior of planetary migration. As a result of these simulations, we find that the initial radial coordinate is a very significant parameter in how the planet interacts with the gas in the disk and the formation of gas gaps.

\vspace{0.8em}
\end{abstract}

\section{INTRODUCTION}

Since the discovery of 51 Pegasi b by \citet{mayor1995} -- the first exoplanet discovered orbiting a solar-like star -- the number of exoplanets has increased tremendously with a vast diversity of extra-solar planetary systems \citep{santerne2011}. Today we have around three thousand exoplanets (Figure 1) detected, and the number keeps getting larger as new methods are developed and the time of observation increases.

The most common methods for planet detection are radial velocity \citep{charbonneau2000, reiners2010}, planetary transit \citep{silvavalio2008}, direct imaging \citep{currie2012}, pulsar timing \citep{wolszczan1992} and gravitational microlensing. In any of these methods, the larger the mass, radius of the exoplanet the easier it is to detect them. Also, the proximity to its host star facilitates the detection since it makes the star more responsive to the planet gravitational attraction.

Planet 51 Pegasi b met all these criteria, thus after its discovery, searches for exoplanets with these features became common. However, an observation bias was established, and a significant number of giant gas exoplanets orbiting near their host stars were discovered. These types of planets are called ``hot Jupiters'' \citep{brucalassi2016} since they are inferred to be physically similar to our Jupiter but with a very short orbital semi-major axes, varying from 0.015 to 0.5 au\footnote{1 astronomical unit = $1.49597870700 \times 10^{11}$ m.} (Figure 2). Nonetheless, the existing models of planetary system formation did not predict this kind of situation.

Before the 90s, our understanding of planetary formation was based on observations of the planets of the Solar System. In particular, the theories of formation were intended to explain the dichotomy of the Solar System, with terrestrial planets in its inner part and gas planets in its outer part. A few years of exoplanet detection brought the need to improve the models and revealed that the formation of our Solar System is not a general rule of planetary system formation.

\begin{figure}[ht]
\centering
\includegraphics[width=0.7\linewidth]{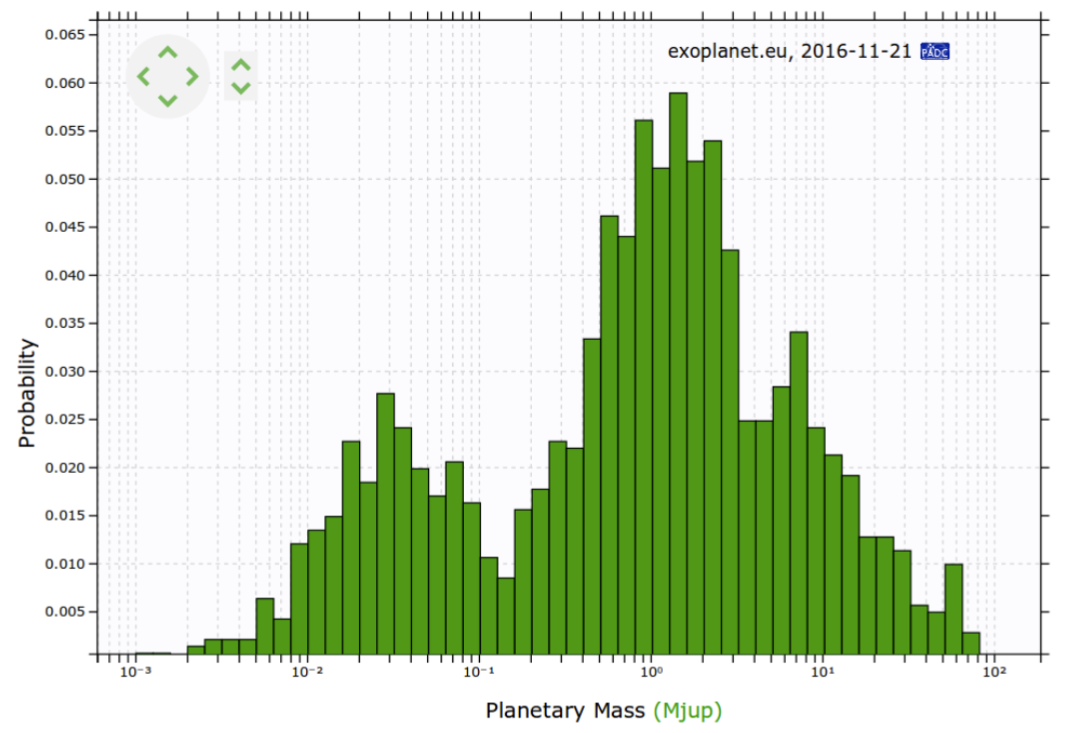}
\caption{Histogram  of  the  planetary  mass  ($M_\text{Jup}=1.898\times10^{27}$~kg).  Note  that  the  majority  of  exoplanets discovered are Jupiter-like planets (data from Feb/2018). Source: \url{http://exoplanet.eu/diagrams/?t=h} on Feb/2018.}
\label{fig:hist}
\end{figure}

\begin{figure}[ht]
\centering
\includegraphics[width=0.4\linewidth]{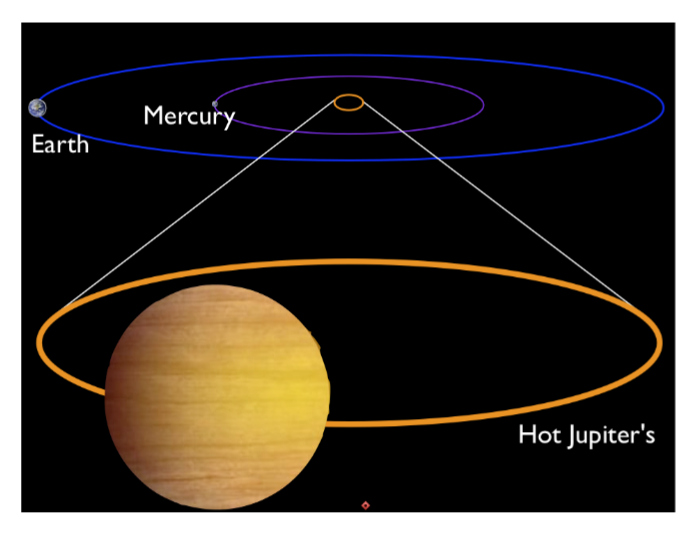}
\caption{Generic example of a hot Jupiter orbit compared to Mercury’s. Source: \url{http://www.stellarplanet.co.uk/2012/11/size-is-everything.html}}
\label{fig:hotJup}
\end{figure}

The generally accepted theory of planet formation from a protoplanetary disk preclude such planets to form so close to their host stars, as there is an insufficient mass at such small radii and the temperature is too high to allow the formation of rocky or icy planetesimals. Therefore, the theories had to be amended to agree with the new discoveries. Currently, the most accepted explanation for the formation of hot Jupiters is planetary migration. This process is believed to occur when a planet in the process of formation interacts with a disk of gas and/or planetesimals, resulting in the modification of its orbital parameters. There are three main types of planetary migration:

\begin{itemize}
\item \textbf{Type I migration.} This type of migration does not produce a gap in the disk and happens relatively fast regarding the million-year lifetime of the gaseous disk. It occurs with small planets with masses less than that of Saturn. Planets can drive spiral density waves in the disk, due to their gravitational potential. In this regime, the disk response to this perturbation is linear. An imbalance occurs in the strength of the interaction with the spirals inside and outside the planet orbit. In most cases, the outer wave exerts a somewhat greater torque on the planet than the internal wave, and this causes the planet to lose orbital angular momentum. The planet then migrates inwards on timescales that are initially short relative to the million-year lifetime of the disk according to \citet{lubow2010}.

\item \textbf{Type II migration.} Planets that are massive enough to open a gap in a gaseous disk, \textit{i.e.}, if the planet is more massive than Saturn, it starts to clear a gap in the disk density distribution. The tidal torque that the planet exerts on the gas transfers angular momentum to the gas exterior of the planet's orbit, and does the opposite interior to the planet, thereby repelling gas from around the orbit and resulting in full gaps in the disk. However, material continues to enter the gap moving the planet and gap inward. The migration rate of this process is lower, making the timescale bigger than the type I and type III regimes as stated by \citet{dangelo2008}, and this is one hypothesis for the formation of most of the hot Jupiters.

\item \textbf{Type III migration.} This type of migration is the newest of the three ones described in the literature \citep{masset2003}. It is also known as ``runaway migration'' due to its high migration rate and very short timescales. In this regime, the planet opens partial gaps in the gas surface density of the disk, so it interacts with large-scale vortices within the disk causing a rapid process of migration. According to \citet{masset2003}, this regime only occurs with planets in the range of 0.04-1.5 $M_{\text{Jup}}$\footnote{Jupiter's mass, $M_{\text{Jup}}$, is $1.898 \times 10^{27}$ kg.} (more common with planets close to 1 $M_{\text{Sat}}$\footnote{Saturn's mass, $M_{\text{Sat}}$, is $5.683 \times 10^{26}$ kg ($\sim 3.34 M_{\text{Jup}}$).}) for an aspect ratio of 5\% (see sections 2.1.1 and 2.1.2). However, \citet{lin2010} showed that this runaway migration could also happen with more massive planets (1-Jupiter-mass planets) in low-viscosity disks.
\end{itemize}

Among the mechanisms proposed to clarify the observed properties of exoplanets, the gravitational interaction between the planets and the protoplanetary disk plays an important role. Understanding the physical processes involved in planet-disk interaction is one of the keys to explaining or deducing the scenarios of planetary systems, including that of the Solar System itself.

\section{METHODOLOGY}

In this study we made numerical simulations using the magneto-hydrodynamics code FARGO3D\footnote{At FARGO3D homepage, \url{http://fargo.in2p3.fr}, there is access to documentation, to download the code and some other features.} described in \citet{benitez2016}, with its similar previous version FARGO used in \citet{lin2010} and in \citet{santos2014}, and its last version used in \citet{fung2017} and \citet{chametla2017}. The selection of this code was because of its excellent reputation in the literature, its easy installation, its simplicity in the initial setup and its outputs in a format of easy reading and analysis.

Five initial scenarios were chosen to run the simulations. The only free parameter was the initial radial position with the aim of verifying the influence of this parameter on the behavior of planetary migration. The simulations were all performed with a 2-Jupiter-mass planet in the protoplanetary disk undergoing gas accretion.

\subsection{The Code FARGO3D}

We ran simulations of a Jupiter-like planet undergoing gas accretion in different orbits of a planetary system formation using the FARGO3D (Fast Advection in Rotating Gaseous Objects - 3D) which is a ``[...] magnetohydrodynamics code developed with special emphasis on the physics of protoplanetary disks and planet-disk interactions, and paralleled with MPI'' according to \citet{benitez2016} and the successor of the FARGO code.

The main features of the code are: solves the equations of hydrodynamics and magnetohydrodynamics on an Eulerian mesh; multidimensional (1D, 2D \& 3D); cartesian, cylindrical and spherical geometries; non inertial reference frames; adiabatic or Isothermal Equation of State (EOS); includes the full viscous stress tensor in the three geometries; simple N-body integrator, for embedded planets.

\subsubsection{Basic Equations}

First, we describe the parameters of the equations solved by the code to perform the simulations.

The planet is localized by its orbital radius $r_p$ and its angular frequency $\Omega_p$. The position of a fluid element on the disk is represented in polar coordinates $r$ and $\varphi$, with origin in the position of the star, while the distance from the fluid element to the orbit of the planet is $x = r - r_p$. The Keplerian frequency is $\Omega_K(r)$, and the orbital frequency of the disk material is $\Omega(r)$. The kinematic viscosity is $\nu$ and the surface density of the disk $\Sigma = \Sigma(r)$, in particular, $\sigma_0$ is the uniform surface density of the undisturbed disk. The mass of the planet is $M_p$, and the mass of the star is $M_*$, the ratio of the mass being defined as $q = M_p / M_*$. The aspect ratio of the disk is $h = H/r$ where $H = H(r)$ is the vertical height of the disk at the position $r$, and finally, $v_r$ and $v_\varphi$ are the radial and azimuthal components of the velocity. Moreover, finally, the hydrodynamic equations:

\begin{itemize}
\item Continuity equation:
\begin{equation}
\frac{\partial \Sigma}{\partial t} + \frac{1}{r} \frac{\partial(r v_r \Sigma)}{\partial r} + \frac{1}{r} \frac{\partial(v_\varphi \Sigma)}{\partial \varphi} = 0
\end{equation}

\item Navier-Stokes equation:
\begin{equation}
\frac{\partial v_r}{\partial t} + v_r \frac{\partial v_r}{\partial r} + \frac{v_\varphi}{r} \frac{\partial v_r}{\partial \varphi} - \frac{v_\varphi^2}{r} = -\frac{1}{\Sigma} \frac{\partial p}{\partial r} - \frac{\partial \Phi}{\partial r} + \frac{f_r}{\Sigma},
\end{equation}
for $v_r$ and
\begin{equation}
\frac{\partial v_\varphi}{\partial t} + v_r \frac{\partial v_\varphi}{\partial r} + \frac{v_\varphi}{r} \frac{\partial v_\varphi}{\partial \varphi} + \frac{v_r v_\varphi}{r} = -\frac{1}{r \Sigma} \frac{\partial p}{\partial \varphi} - \frac{1}{r} \frac{\partial \Phi}{\partial \varphi} + \frac{f_\varphi}{\Sigma},
\end{equation}
for $v_\varphi$, where $p$ is the vertically integrated pressure, $f_r$ and $f_\varphi$ are respectively the radial and azimuthal components of the viscous force per surface unit, and $\Phi$ is the gravitational potential.

\item The equation of state:
\begin{equation}
p = c_s^2 \Sigma,
\end{equation}
for an isothermal gas with sound speed $c_s$.

\item Gravitational potential:
\begin{equation}
\begin{split}
\Phi(r, \varphi) = & -\frac{G M_*}{r} - \frac{G M_p}{\sqrt{r^2 + r_p^2 - 2 r r_p \cos(\varphi) + \epsilon}} + \frac{G M_p}{r_p^2} r \cos(\varphi) \\
& + r \int_{\text{disk}} \frac{G \Sigma(r', \varphi')}{r'^2} \cos(\varphi - \varphi')\,r'\,dr'\,d\varphi'
\end{split}
\end{equation}
where protoplanet potential smoothness factor is $\epsilon = 0.6$.

\item Viscous force tensors:
\begin{align}
f_r &= \frac{1}{r}\frac{\partial(r \tau_{rr})}{\partial r} + \frac{1}{r}\frac{\partial \tau_{r\varphi}}{\partial \varphi} - \frac{\tau_{\varphi\varphi}}{r}, \\
f_\varphi &= \frac{1}{r}\frac{\partial(r \tau_{\varphi r})}{\partial r} + \frac{1}{r}\frac{\partial \tau_{\varphi\varphi}}{\partial \varphi} - \frac{\tau_{r\varphi}}{r},
\end{align}
where the viscous stress tensors are
\begin{align}
\tau_{rr} &= 2\eta D_{rr} - \frac{2}{3}\eta \nabla \vec{v}, \\
\tau_{\varphi\varphi} &= 2\eta D_{\varphi\varphi} - \frac{2}{3}\eta \nabla \vec{v}, \\
\tau_{r\varphi} &= \tau_{\varphi r} = 2\eta D_{r\varphi},
\end{align}
and further
\begin{align}
D_{rr} &= \frac{\partial v_r}{\partial r}, \\
D_{\varphi\varphi} &= \frac{1}{r}\frac{\partial v_\varphi}{\partial \varphi}, \\
D_{r\varphi} &= \frac{1}{2}\left[ r \frac{\partial}{\partial r}\left(\frac{v_\varphi}{r}\right) + \frac{1}{r}\frac{\partial v_r}{\partial \varphi} \right],
\end{align}
where $\eta = \Sigma \nu$ is the integrated dynamic viscosity coefficient.
\end{itemize}

\subsubsection{Boundary Conditions and Other Characteristics of the Simulation}

The simulation was made in a 2D-mesh, where the horizontal directions represent the azimuth (or angular) axis and the vertical direction represents the radial axis. Although it is a rectangular mesh, it was set as cylindrical geometry option so that the code does the 2D calculations considering the disk thickness, $H$, obtained by the aspect ratio, $h_0 = H/r$, where $r$ is the initial radius of the planet.

We did not consider the magnetic field in this study, so the code does not solve the magnetohydrodynamic equations, but only the hydrodynamic equations that were described in section 2.1.1. Two other characteristics are that the disc is made of an isothermal gas, as indicated in equation 4, and the initial density surface is constant (see section~\ref{sec:input}).

\subsubsection{Input Parameters} \label{sec:input}

There are two kinds of input: the planetary parameters and the orbital parameters. The planetary ones are:
\begin{itemize}
\item number of planets, $N_p = 1$;
\item ratio of the planet mass and the star mass, $q = 0.002$;
\item accretion rate, $Z_{\text{accr}} = 0$, which means that the planet does not have its mass increased over time;
\item indication of whether the planet will or will not feel (1 or 0) the gravitational potential of the accretion disk: 1;
\item indication of whether the planet will or will not feel (1 or 0) the gravitational potential of other planets: 0 since we simulated a single planet.
\end{itemize}

The orbital parameters characterize the physical aspects of the disk. There are a large number of parameters that can be changed, among them are:
\begin{itemize}
\item aspect ratio, $h_0 = H/r = 0.05$, where $H$ is the thickness (z-axis) of the disk;
\item initial surface density, $\sigma_0 = 6.3662 \times 10^{-4}$, that is multiplied by $M_* / r_0^2$ [kg/m$^2$];
\item grid parameters such as the number of radial and azimuthal zones, $N_r = 256$ and $N_a = 512$;
\item inner and outer grid boundaries, $R_{\text{in}} = 0.1$ au and $R_{\text{out}} = 15$ au;
\item number of iterations, $N_{\text{tot}} = 1000$, and time step
\begin{equation}
dt = 2\pi \sqrt{\frac{r_0^3}{G M_\odot}}
\end{equation}
\end{itemize}

We also changed the kinematic viscosity of the disk, which is given by
\begin{equation}
\nu = \nu_0 \times 10^{-5} r_0^2 \Omega(r).
\end{equation}
The value of $\nu_0$ was set to 0.1 to induce a runaway migration even with a planet more massive than Jupiter. This way we could test the most recent type of migration described, with a different viscosity and a different planet mass, and make the computational process faster since this regime is physically faster than type II migration.

\subsection{The simulation}

\subsubsection{Setup}

In total five simulations were performed. The difference between them was the initial position of the planet given by its radial coordinate $r_0$ which were 0.5, 1.2, 3.2, 5.2 and 11.2 au. These initial positions were chosen because they are similar to the distances of Earth, Mars, Jupiter and Saturn distances to the Sun. From the $r_0$ values and using Eq. 14, we can calculate the time steps for each simulation, these are listed in Table 1.

\begin{table}[ht]
\centering
\caption{Time steps for each $r_0$.}
\label{tab:1}
\vspace{0.4em}
\begin{tabular}{lccccc}
\toprule
$r_0$ (au) & 0.5 & 1.2 & 3.2 & 5.2 & 11.2 \\
$dt$ (yr) & 0.3536 & 1.315 & 5.725 & 11.86 & 37.48 \\
\bottomrule
\end{tabular}
\end{table}

The mass of the planet used was 2.0 $M_{\text{Jup}}$, \textit{i.e.}, mass ratio $q = 0.002$ since the star was considered a Sun-like star ($M_\star = 1~M_\odot$ \footnote{Solar mass, $M_\odot = 1.9891 \times 10^{30}$ kg}). The rest of the parameters were kept the same in all simulations.

\section{RESULTS AND DISCUSSION}

After the simulations were done, we analyze the gas surface density over the disk and the variations of the planet radial coordinate $r$ as a function of the orbit (or time). In Figs. 3, 4, 5, 6 and 7 we can observe the gas surface density distribution along the disk for $r_0$ values of 0.5, 1.2, 3.2, 5.2 and 11.2 au, respectively. These figures are divided into two plots: output number 50 (left) and output number 1000 (right), which are respectively the instants $t_i = dt \times \text{output number}$. The text and dotted line in blue represent the initial coordinate $r_0$ and the text and dotted line in green represent the final coordinate $r_f$. The color bar indicates a color scale that shows the values of gas surface density in $\log_{10}(\sigma_0 M_* / r_0^2)$.

Besides, it should be noted that the $R$ limits in the figures are not the same as the simulated grid (from 0.1 au to 15.2 au). This was made to amplify the details in the inner part. In Figure 3 the outer boundary is 1 au; in Figs. 4 and 5 the outer boundary is 5 au; in Figure 6 is 10 au; whereas in Figure 7 it is 15 au.

\begin{figure}[ht]
\centering
\includegraphics[width=0.75\linewidth]{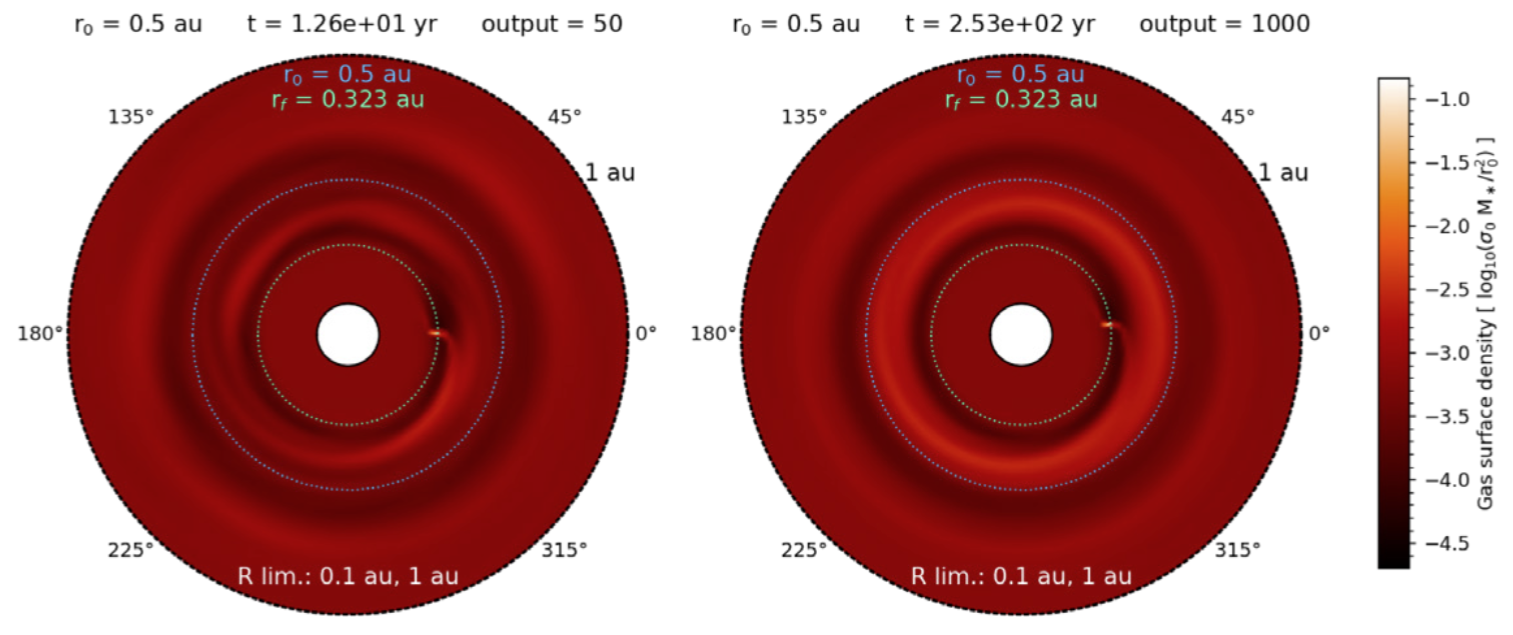}
\caption{2D projection of gas surface density for $r_0 = 0.5$ au. Left panel: $t = 1.26 \times 10^1$~yr (output = 50). Right panel: $t = 2.53 \times 10^2$~yr (output~=~1000)}
\label{fig:3}
\end{figure}

\begin{figure}[ht]
\centering
\includegraphics[width=0.75\linewidth]{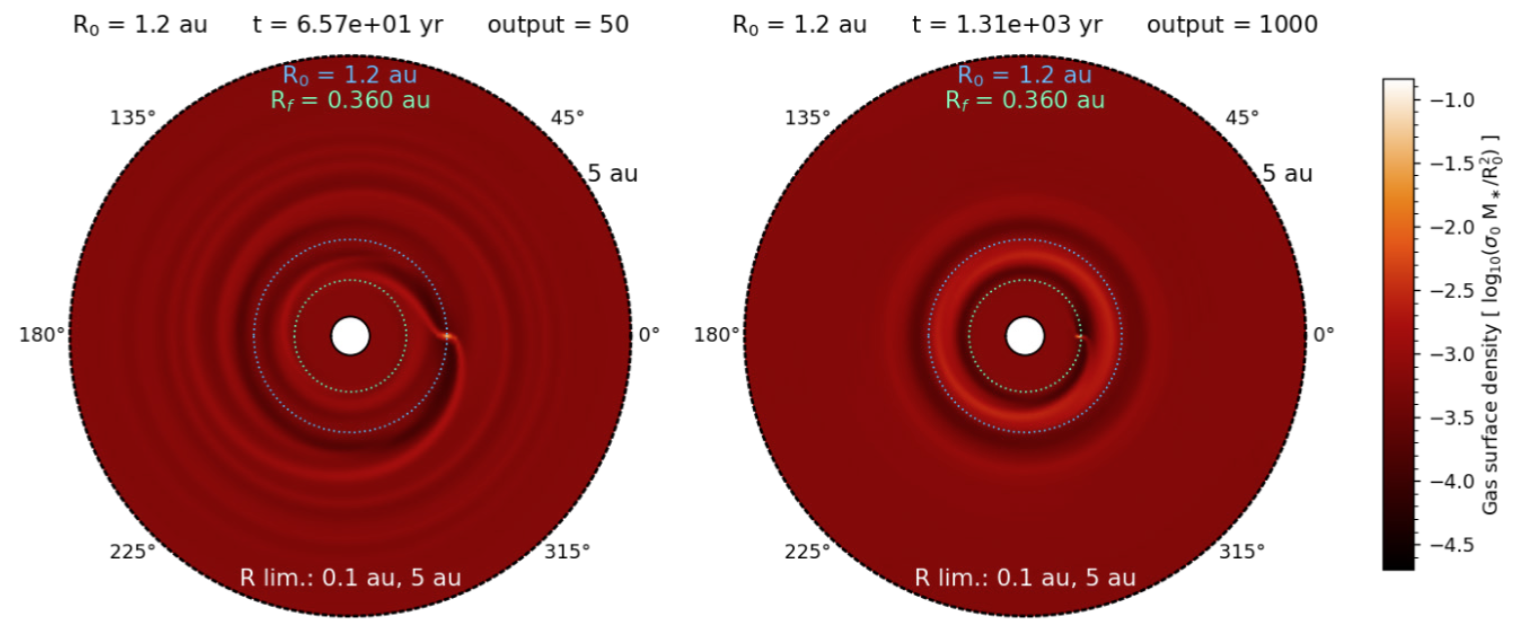}
\caption{2D projection of gas surface density for $r_0 = 1.2$ au. Left panel: $t = 6.57 \times 10^1$~yr (output = 50). Right panel: $t = 1.31 \times 10^3$~yr (output~=~1000).}
\label{fig:4}
\end{figure}

\begin{figure}[ht]
\centering
\includegraphics[width=0.75\linewidth]{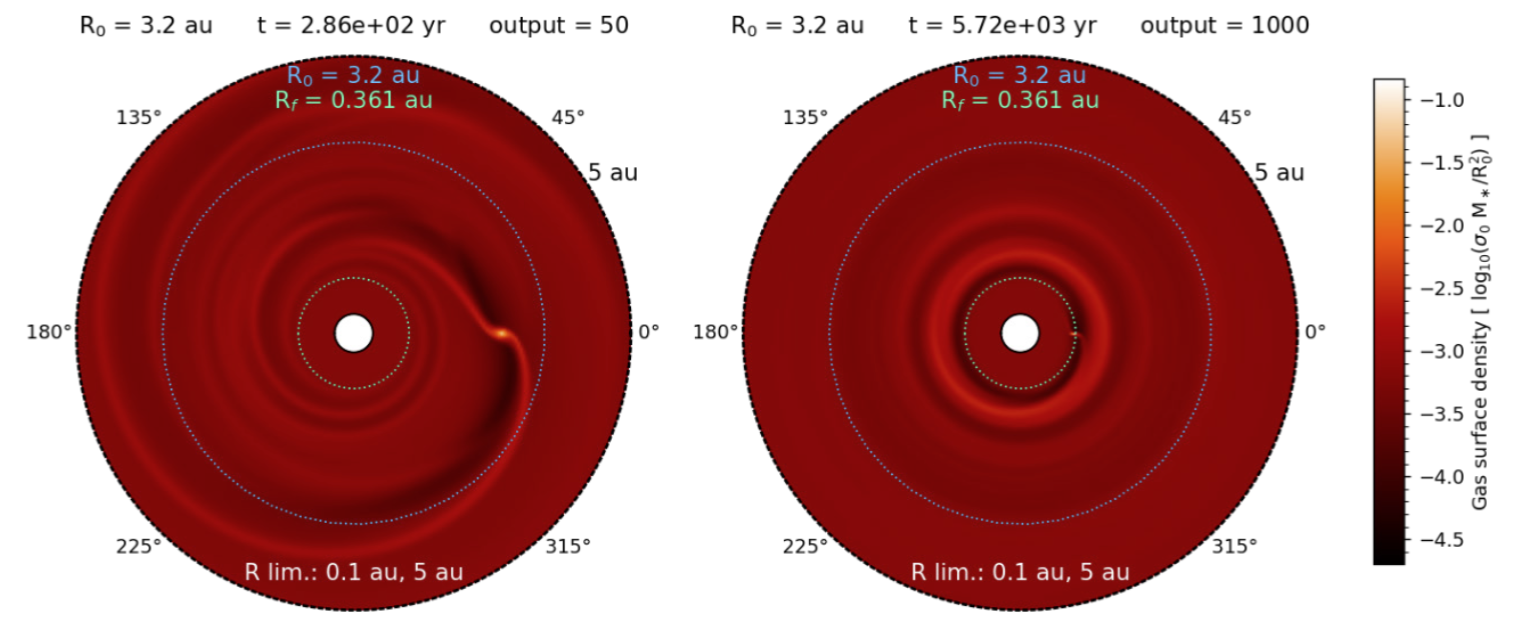}
\caption{2D projection of gas surface density for $r_0 = 3.2$ au. Left panel: $t = 2.86 \times 10^2$~yr (output = 50). Right panel: $t = 5.72 \times 10^3$~yr (output~=~1000).}
\label{fig:5}
\end{figure}

\begin{figure}[ht]
\centering
\includegraphics[width=0.75\linewidth]{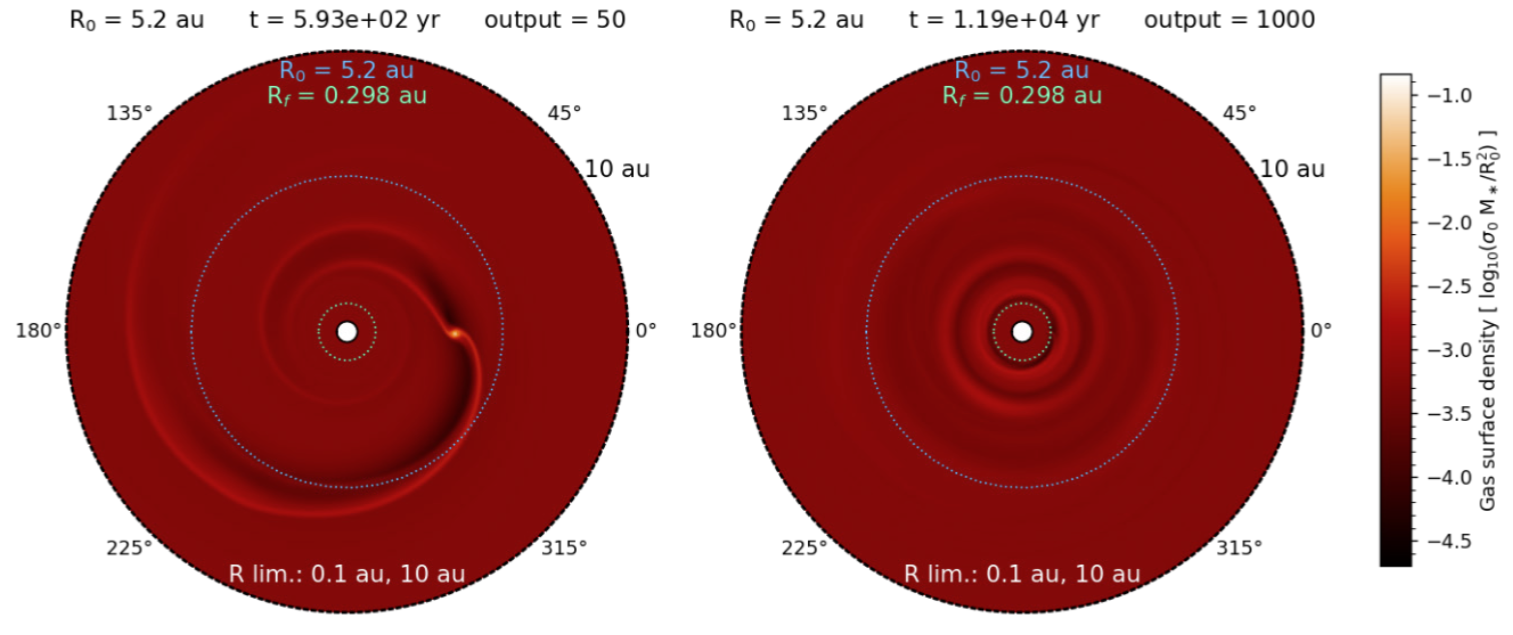}
\caption{2D projection of gas surface density for $r_0 = 5.2$ au. Left panel: $t = 5.93 \times 10^2$~yr (output = 50). Right panel: $t = 1.19 \times 10^4$~yr (output~=~1000).}
\label{fig:6}
\end{figure}

\begin{figure}[ht]
\centering
\includegraphics[width=0.75\linewidth]{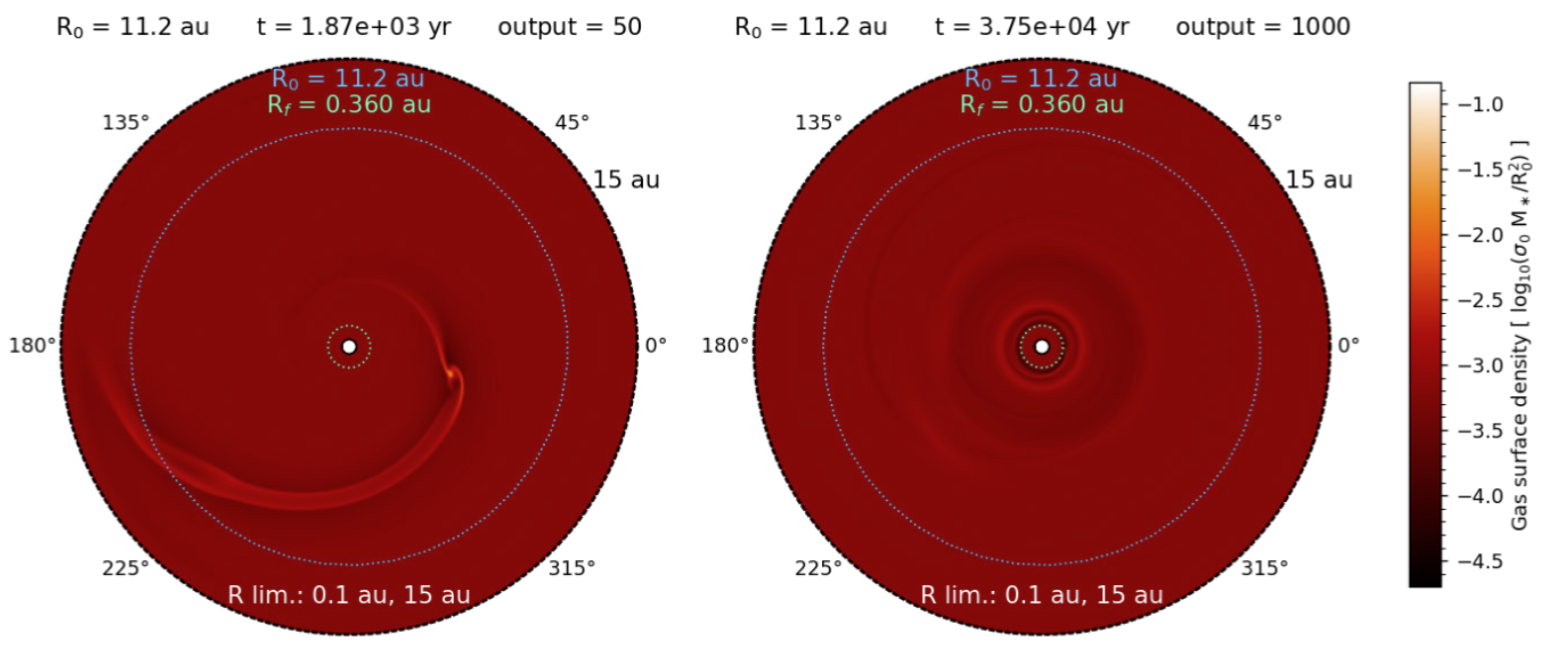}
\caption{2D projection of gas surface density for $r_0 = 11.2$ au. Left panel: $t = 1.87 \times 10^3$~yr (output = 50). Right panel: $t = 3.75 \times 10^4$~yr (output~=~1000).}
\label{fig:7}
\end{figure}

All simulations resulted in the formation of gaps in the accretion disks. Moreover, the final coordinate of the planetary migration was $r_f < 0.5$ au, classifying these planets as hot Jupiters. Three of them had their final coordinates stabilized at approximately 0.36 au. In the first simulation ($r_0 = 0.5$ au) the planet had its radial position stabilized at 0.328 au, while the planet initially at $r_0 = 5.2$ au, stabilized at 0.298 au. The total variations ($r_0 - r_f$) are listed in Table 2.

\begin{table}[ht]
\centering
\caption{Radial Coordinate Results.}
\label{tab:2}
\vspace{0.4em}
\begin{tabular}{lccccc}
\toprule
$r_0$ (au) & 0.5 & 1.2 & 3.2 & 5.2 & 11.2 \\
$r_{\min}$ (au) & 0.324 & 0.350 & 0.341 & 0.243 & 0.350 \\
$r_0 - r_f$ (au) & 0.177 & 0.840 & 2.84 & 4.90 & 10.8 \\
$r_f$ (au) & 0.328 & 0.360 & 0.361 & 0.298 & 0.360 \\
\bottomrule
\end{tabular}
\end{table}

\begin{figure}[ht]
\centering
\includegraphics[width=0.65\linewidth]{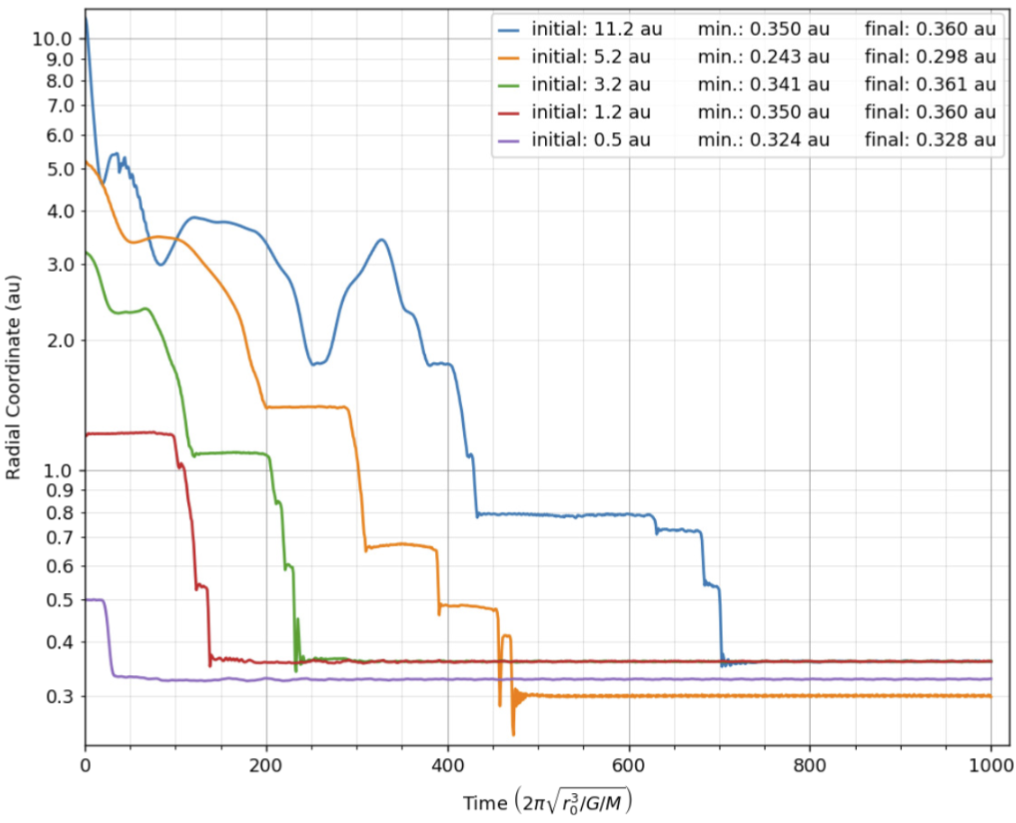}
\caption{Planet radial coordinates versus time.}
\label{fig:8}
\end{figure}

Figure 8 displays the position of the planet in time represented by the radial coordinate as a function of the output number, where each color represents one of the 5 simulations. The presence of plateaus as seen on the figure corresponds to epochs of radial stabilization. They are associated with the formation of partial gaps, \textit{i.e.}, that does not surround the entire disk. When the gap is filled, it means that the planet orbital radius is stabilized.

In all simulations, except for $r_0 = 0.5$ au, there are plateaus close to 1.0 au. The mean radii of them are 1.22, 1.35, 1.40 and 0.787 au for each $r_0$ respectively from 1.2 to 11.2 au. Another feature of the output of the simulations is that, as the initial radial position increases, there are more pronounced variations in radius.

\section{CONCLUSIONS}

In this study, numerical simulations of a planet undergoing gas interaction were performed with the FARGO3D code, which solves the hydrodynamics equations in a 2D mesh grid. A total of five simulations were made with only one free parameter -- the initial radial position -- to verify its influence on the behavior of the interaction between the protoplanetary disk and the planet.

As a result, the planet initial radial coordinate, $r_0$, affects the way the planet interacts with the gas in the disk. All outcomes showed the planet to stabilize close to 0.34 au. The higher the radial distance of the planet at the beginning of the simulation, the higher the total radial variation due to the formation of gas gaps. These results are impressive because they show that independently of where the protoplanet is initially formed, its final position will be the same, about 0.34 au in the case of these simulations.

All of them described a type III migration that is, the sharp variations in radial coordinates with the formation of partial gaps in the disk and was very fast (from years to hundreds of years). As was stated in the description of type III migration, this occurs only in cases of the planet with a mass of the order of 1 $M_{\text{Sat}}$. However, as \citet{lin2010} have shown, 1-Jupiter-mass planets can be subjected to this regime if the gas disk is of low viscosity. In our case, using a low-viscosity disk ($\nu_0 = 0.1$, from Eq. 15) and a planet two times more massive, we oberved a type III migration.

Another interesting outcome of these simulations is the temporary stabilization around 1 au since this is the current location of Earth and is within the habitable zone of our Sun. It could be related to the density of the disk and the mass of the protoplanet, which would determine the torque interaction of them over time \citep{tanaka2002}. Also, other parameters like stellar mass could dictate variations on those results. Future work of varying other initial parameters will be crucial to determining which ones define the planet final position.




\vspace{1em}
\noindent \textbf{Contact:} Fabian Marcel Menezes \href{mailto:menezes.astroph@gmail.com}{menezes.astroph@gmail.com} \\
\textbf{Process:} Submission on February 2018. Acceptance on June 2018.

\appendix
\section*{APPENDIX}

The FARGO3D code runs in parallel with the MPI procedure. A given calculation can be done by multiple CPUs simultaneously. The executions of the simulations are computationally expensive. To do this, we used the ``Wintermute'' cluster, installed in the Center for Radio Astronomy and Astrophysics at Mackenzie (CRAAM), at the Mackenzie Presbyterian University.

As is explained by \citet{santos2014}, the Wintermute cluster is a H8DGT-HIBQF model, with 500 GB SATA II hard drive and four motherboards. Each motherboard consists of two AMD Magny-Cours 12C processors, 64-bit, 1.7 GHz clock and twelve cores. Also, each one has 12 GB of DDR3 ECC RAM. The motherboards were installed in a twin-type enclosure with a redundant source of 1,400 W. In total, Wintermute has 96 cores and 48 GB of RAM. The connection is via two 1 GB/s ports to each motherboard, and the nodes communicate with each other via a 3COM 24-port 1 GB/s switch and 4 SFP ports. In the four nodes, the Debian operating system was installed along with the OpenMPI and MPICH distributed processing packages.

It took about 6 hours for each simulation to be completed.

\end{document}